\documentclass[aps,prd,amsfonts,amssymb,preprint,eqsecnum,nofootinbib]
{revtex4} 
\usepackage{graphics,epsfig}
\begin{document}
\preprint{WM-06-104}
%
\title{\vspace*{0.5in} Unusual High-Energy Phenomenology of \\
Lorentz-Invariant Noncommutative Field Theories
\vskip 0.1in}
\author{Christopher D. Carone}\email[]{carone@physics.wm.edu}
\author{Herry J. Kwee}\email[]{hxjohn@wm.edu}
\affiliation{Particle Theory Group, Department of Physics,
College of William and Mary, Williamsburg, VA 23187-8795}
\date{March 2006}
\begin{abstract}
It has been suggested that one may construct a Lorentz-invariant noncommutative field theory
by extending the coordinate algebra to additional, fictitious coordinates that transform
nontrivially under the Lorentz group.  Integration over these coordinates in the action produces
a four-dimensional effective theory with Lorentz invariance intact.  Previous applications of this
approach, in particular to a specific construction of noncommutative QED, have been studied only in a
low-momentum approximation.  Here we discuss Lorentz-invariant field theories in which the relevant physics
can be studied without requiring an expansion in the inverse scale of noncommutativity. Qualitatively, we
find that tree-level scattering cross sections are dramatically suppressed as the center-of-mass energy
exceeds the scale of noncommutativity, that cross sections that are isotropic in the commutative limit can
develop a pronounced angular dependence, and that nonrelativistic potentials (for example, the Coloumb
potential) become nonsingular at the origin.  We consider a number of processes in noncommutative QED
that may be studied at a future linear collider. We also give an example of scattering via a four-fermion
operator in which the noncommutative modifications of the interaction can unitarize the tree-level amplitude,
without requiring any other new physics in the ultraviolet.
\end{abstract}
\maketitle

\section{Introduction}\label{sec:intro}
Noncommutative field theories have provoked considerable interest from both the
formal~\cite{formal} and phenomenological~\cite{phenom} perspectives.  In their simplest construction, such
theories promote spacetime coordinates to operators (indicated below by a hat) that
satisfy a nontrivial commutation relation
\begin{equation}
[ \hat{x}^\mu\,,\,\hat{x}^\nu] = i \, \theta^{\mu\nu} \,\,\,,
\label{eq:opcom}
\end{equation}
where $\theta^{\mu\nu}$ is an antisymmetric, constant matrix. The noncommutative
multiplication of operators is represented via function of ordinary, commuting
coordinates by means of a modified multiplication rule, the star product
\begin{equation}
f(x) \star g(x) = f(x) \, e^{\frac{i}{2}\stackrel{\leftarrow}{\partial} \cdot \theta \cdot
\stackrel{\rightarrow}{\partial}} g(x)  \,\,\,.
\label{eq:starprod}
\end{equation}
For example, it is easy to show that the commuting coordinates $x^\mu$ satisfy the star
commutation relation
\begin{equation}
[ x^\mu\,\stackrel{\star}{,}\, x^\nu]=i \,\theta^{\mu\nu} \,\,\,,
\end{equation}
mimicking the behavior of the operators in Eq.~(\ref{eq:opcom}).  Noncommutative fields
theories are related to their commutative cousins via the promotion of ordinary
multiplication to star multiplication, and the imposition of restrictions on the
form of the interactions as needed to maintain the desired local symmetries
of the theory~\cite{ncgauge}.

The effects of noncommutative spacetime on rare processes~\cite{rare}, collider
signals~\cite{collider}, astrophysics~\cite{astro}, and cosmology~\cite{cosmo} have been
discussed in the literature over the past few years.  One notable feature of noncommutative
field theories based on Eq.~(\ref{eq:opcom}) is that they violate Lorentz invariance; despite
its appearance, the matrix $\theta^{\mu\nu}$ is a singlet under the Lorentz group, and
defines preferred the directions $\epsilon^{ijk} \theta^{jk}$ and $\theta^{0i}$.  This has
led to the observation that collider signatures of noncommutative physics should present a diurnal
variation, as the earth's orientation relative these preferred directions varies
over time~\cite{diurnal}. On the other hand, noncommutative field theories must contend with very
stringent bounds from processes measured precisely at low energies~\cite{lowe}.  For example, in
almost every construction of Lorentz-violating noncommutative QED, there are
contributions to the Lamb shift which force the scale of new physics to be above
$\sim 10$~TeV~\cite{alank}.  Moreover, in the simplest formulation of noncommutative gauge theories (for
example, the version of noncommutative QED first introduced by Hayakawa~\cite{haya}), one may show that
operators induced at the one-loop level present even more stringent bounds on the noncommutative scale,
whether the loops are evaluated with a simple cut-off or a gauge-invariant
regulator~\cite{abdg,CCLsusy,ferrari}.  Noncommutative gauge theories based on the enveloping algebra
approach~\cite{jurco} contribute to the Lamb shift at tree-level, and are thus subject to the $10$~TeV
bound. These theories may also be constrained significantly via loop-induced operators~\cite{CCLqcd}, though
in this case the situation is less clear~\cite{calmet}.  However one views the severity of these bounds, it
is clear in any construction of noncommutative field theories based on Eq.~(\ref{eq:opcom}), the matrix
$\theta^{\mu\nu}$ will appear in the low-energy effective Lagrangian as a Lorentz-violating spurion.  By
contrast, all experimental searches for the violation of Lorentz invariance have yielded negative results to
date~\cite{lvexp,morelv}.

These observations provide some motivation for considering whether noncommutative theories can
be constructed which preserve Lorentz invariance {\em ab initio}\footnote{An alternative
approach is to construct a noncommutative theory that preserves a deformation of the Lorentz
group~\cite{deform}. For other discussions of Lorentz symmetry in noncommutative theories, see
Refs.~\cite{oapp1,oapp2}}.  In this paper, we will revisit
a proposal made in Ref.~\cite{CCZ} (and studied in Refs.~\cite{m1,m2,m3,CKN,hag}) which approaches this
problem by promoting $\theta^{\mu\nu}$ to an operator in Eq.~(\ref{eq:opcom}),
\begin{equation}
[ \hat{x}^\mu\,,\,\hat{x}^\nu] = i \, \hat{\theta}^{\mu\nu} \,\,\,,
\label{eq:opcomn}
\end{equation}
where $\hat{\theta}^{\mu\nu}$ transforms as a tensor under the Lorentz group. In addition, one assumes
the simplest possibility $[\hat{x}^\mu,\hat{\theta}^{\alpha\beta}]=
[\hat{\theta}^{\mu\nu},\hat{\theta}^{\alpha\beta}]=0$. The star product for this algebra takes the same
form as Eq.~(\ref{eq:starprod}), except that it acts, in general, on functions of six additional commuting
coordinates $\theta^{\mu\nu}$
\begin{equation}
f(x,\theta) \star g(x,\theta) = f(x,\theta) \, e^{\frac{i}{2}\stackrel{\leftarrow}{\partial} \cdot
\theta \cdot \stackrel{\rightarrow}{\partial}} g(x,\theta)  \,\,\,.
\label{eq:starprodn}
\end{equation}
As argued in Ref.~\cite{CCZ}, the action will involve integration over the additional coordinates
\begin{equation}
{\cal L}_{\mbox{eff}} = \int d^6\theta \, W(\theta) \, {\cal L}(x^\rho, \theta^{\mu\nu})
\label{eq:effL}
\end{equation}
so that ${\cal L}_{\mbox{eff}}$ depends only on ordinary coordinates and has four-dimensional Lorentz
invariance preserved. The weighting function $W(\theta)$ is a Lorentz-invariant function of $\theta^{\mu\nu}$
that has finite volume and hence drops off at large values of the argument. In the mapping from the space of
operators to c-number functions, it was argued in Ref.~\cite{CCZ},  that the $d^6\theta\, W(\theta)$ integration
is an appropriate mapping of the operator trace, which is reasonable to believe plays a role in the construction
of an action.  In the present context, it is only important to note that $W(\theta)$ determines the scale
of the new physics.  For example, the choice $W(\theta)=\delta^{(6)}(\theta)$ pushes the scale of
noncommutativity to infinity.

Two question deserve immediate comment:  (1) Is there anything wrong with treating
$\theta^{\mu\nu}$ as a set of fictitious coordinates with nontrivial Lorentz transformation properties
and (2) where do ordinary quantum fields live within the right-hand-side of Eq.~(\ref{eq:effL})?

On the first question, it is useful to consider the Grassman parameters $\theta$ and $\bar{\theta}$
of ordinary superspace.  These may be thought of as fictitious coordinates that transform nontrivially
under the Lorentz group and are integrated in the action to produce a Lagrangian that is invariant under
global supersymmetry transformations.  Extending the coordinate algebra by the bosonic coordinates
$\theta^{\mu\nu}$ in Eq.~(\ref{eq:effL}) is neither more nor less justified mathematically then
extending ordinary space to superspace.  On the other hand, Doplicher, Fredenhaeghen and Roberts~\cite{DFR}
have argued on general grounds that theories of quantum gravity should lead at low energies to
spacetime noncommutativity described by a commutation relation of precisely the same form as
Eq.~(\ref{eq:opcomn}), one in which the right-hand-side is an operator that transforms nontrivially under
the Lorentz group so that the result remains covariant\footnote{One can also understand
the algebra of interest to us here as a contraction of the one proposed by Snyder~\cite{snyder}.  See
Ref.~\cite{CCZ} for details.}.

On the second question, we may assume that the Lagrangian ${\cal L}$ is a function of fields $\Phi$
that are themselves functions of $x^\mu$ and $\theta^{\mu\nu}$, with all multiplications between fields
promoted to star multiplications.  In the supersymmetric analogy, superfields are functions of all the
superspace coordinates, while ordinary fields appear in the coefficients of an expansion of the
superfields in $\theta$ and $\bar{\theta}$. It was shown in Ref.~\cite{CCZ} that an analagous statement
can be made for the fields $\Phi(x^\rho,\theta^{\mu\nu})$
\begin{equation}
\Phi(x,\theta)= \phi(x) + \theta^{\mu\nu}\,\Phi^{(1)}_{\mu\nu}(x) +
\theta^{\mu\nu}\theta^{\eta\sigma}\,\Phi^{(2)}_{\mu\nu\eta\sigma}(x)+\cdots \,\,\,,
\label{eq:bigexp}
\end{equation}
where the $\Phi^{(n)}$ are functions of the ordinary quantum field $\phi(x)$.  (Of course, unlike the
supersymmetric example, the expansion here does not truncate.) In Ref.~\cite{CCZ}, it was shown that the
constraint of gauge invariance allows one to determine the $\Phi^{(n)}$ as functions of the quantum field
$\phi$ and the gauge fields in the theory.  This allowed for the construction of a Lorentz-invariant version
of noncommutative QED, one that was valid for fields of arbitrary charge.

One of limitation of the models discussed in Refs.~\cite{CCZ,CKN} is that they could only be evaluated
as an expansion in $\theta$ (or more accurately, after the $d^6\theta$ integration, as an expansion in
momentum divided by the typical noncommutative energy scale).  The phenomenological study of Lorentz-invariant
noncommutative QED in Ref.~\cite{CKN} focused, therefore, on the effects of the new higher-dimension operators
and presented bounds on a natural definition of the noncommutative scale
\begin{equation}
\Lambda_{NC}^{-4} = \frac{1}{12} \int d^6 \theta  \, \theta^{\mu \nu} \theta_{\mu \nu} \, W(\theta) \,\,\,.
\label{eq:lncdef}
\end{equation}
An advantage of this approach is that the form of the weighting function $W(\theta)$ need not be specified, at
least for lowest-order processes.  Comparison of Bhabha scattering, dilepton and diphoton production to LEP data
led the authors of Ref.~\cite{CKN} to the bound
\begin{equation}
\Lambda_{NC} > 160\mbox{ GeV} \,\,\,\,\,\,\mbox{95\% C.L.}
\label{eq:cknbound}
\end{equation}

The purpose of the present work is to consider toy and more realistic examples where
Lorentz-invariant noncommutative theories can be evaluated without a low-momentum expansion.   The toy models
will consist of the Yukawa and $\phi^4$ theories.  Since the expansion in Eq.~(\ref{eq:bigexp}) is not
restricted by gauge invariance in these theories, there is no inconsistency in making the simplest choice
that might allow us to obtain results in closed form
\begin{equation}
\Phi(x,\theta) = \phi(x) \,\,\,.
\end{equation}
The more realistic theory that we will study is the original formulation of noncommutative
QED~\cite{haya}, which also does not require a $\theta$-expansion for the fields, at the expense of introducing
a  restriction that the matter fields have charges $\pm 1$ or $0$.  This will allow us to revisit the purely
leptonic processes considered in the low-momentum limit of the alternative formulation discussed in
Ref.~\cite{CKN}.  Finally, to avoid an expansion in $\Lambda_{NC}^{-1}$, we will have to choose a tractable
form for the weighting function $W(\theta)$.  Neither Ref.~\cite{CCZ} or \cite{m1,m2,m3,CKN,hag}
ever specify $W(\theta)$, or show that a function with the desired properties exists.  We will find the
simplest form for this function, based on the quadratic invariants of $\theta^{\mu\nu}$, that will allow for
explicit calculations at arbitrary external momenta.

Our paper is organized as follows.  In Section~\ref{sec:two}, we show that suitable weighting
functions exist, and we determine a simple form that will be used in evaluating the models that follow. In
Section~\ref{sec:three} we will look at the behavior of tree-level scattering cross sections in $\phi^4$ and
Yukawa theory.  We will see that cross sections experience a notable suppression as the center-of-mass
energy exceeds the typical scale of the noncommutative interactions and that isotropic, tree-level
processes in the commutative limit can develop a marked angular dependence.  In Section~\ref{sec:four} we will
extend our analysis to noncommutative QED, focusing in particular on Bhabha scattering, dilepton
and diphoton production.  We will also show that the modified momentum dependence of the tree-level
fermion-photon vertex removes the singularity at the origin of the Coloumb potential. In Section~\ref{sec:five},
we show that a simple theory that violates tree-level unitarity in the commutative limit can be be unitarized
by the noncommutative modification to the vertex.  In Section~\ref{sec:six}, we summarize our conclusions.

\section{The Weighting Function}\label{sec:two}

In this section, we find a suitable explicit form for the weighting function $W(\theta)$ in
Eq.~(\ref{eq:effL}).  In particular, we would like $W(\theta)$ to be Lorentz invariant and
normalizable~\cite{CCZ}.  If the new coordinates carried only a single Lorentz index ({\em i.e.},
if they transformed like the ordinary coordinates $x^\mu$) it would not be possible to find such
a function:  All Lorentz-invariant functions of $x^\mu$ are necessarily a function of $x^\mu x_\mu$,
which vanishes along its light cone.  Thus, integration over all four-volume would diverge at
large $x^\mu$ in the direction where $x^\mu x_\mu=0$.

As pointed out by Kase, {\em et al.}, the same is true of weighting functions $W(\theta)$ that
are only functions of $\theta^{\mu\nu} \theta_{\mu\nu}$~\cite{m2}.  One approach, taken
in Ref.~\cite{m2}, is to use a very manageable function, for example a Gaussian in
$\theta^{\mu\nu} \theta_{\mu\nu}$, and regulate away the unwanted divergence.  Here we will show that
simple weighting functions $W(\theta)$ exist that have finite volume and are not so complicated that
they render explicit calculations intractable.  In contrast to a coordinate with a single Lorentz
index, $\theta^{\mu\nu}$ allows for the construction of an infinite number of Lorentz invariant combinations,
for example
\begin{equation}
\theta^{\mu\nu} \theta_{\mu\nu}, \,\,\,\,\,
\theta^{\mu\nu} \theta_{\nu\rho} \theta^{\rho\,\alpha} \theta_{\alpha\mu}, \,\,\,\,\,
\theta^{\mu\nu} \theta_{\nu\rho} \theta^{\rho\,\alpha} \theta_{\alpha\beta} \theta^{\beta \delta}
\theta_{\delta\mu}, \,\,\,\,\, \mbox{etc.}
\end{equation}
In general, the directions in which these invariants are vanishing differ from one to another.  Thus,
we seek a function that drops off at large values of all of its arguments and that depends on a sufficient
number of these invariants so that no flat directions remain.

The simplest possibility is to work only with the two available quadratic invariants
\begin{equation}
\frac{1}{2}\theta_{\mu\nu}\theta^{\mu\nu} \,\,\,\,\,\mbox{ and }\,\,\,\,\,
\frac{1}{8}\epsilon_{\mu\nu\alpha\beta}\theta^{\mu\nu}\theta^{\alpha\beta} \,\,\,,
\end{equation}
where the numerical factors are introduced for later convenience.  To make integration as easy
as possible, we choose to work with an exponential form for $W(\theta)$.  One might begin by considering
the function
\begin{equation}
W_{bc}(\theta) \equiv {\rm exp}(-\frac{c}{2}|\theta_{\mu\nu}\theta^{\mu\nu}|) \,
{\rm exp}(-\frac{b}{8}|\epsilon_{\mu\nu\alpha\beta}\theta^{\mu\nu}\theta^{\alpha
\beta}|)\,\,\,,
\end{equation}
where $b$ and $c$ are constants, and the absolute value symbols are introduced to assure that there are
no directions in which the weighting function blows up at infinity.  Unfortunately, the function
$W_{bc}(\theta)$ is not adequate since integration over all $\theta^{\mu\nu}$ leads to a divergent result.
However, by studying this divergence we will see that there is a simple way to avoid it.

Let us parameterize the components of $\theta^{\mu\nu}$ as follows:
\begin{equation}
\theta^{\mu\nu} = \left(
\begin{array}{cccc}
0 & x & y & z \\
-x & 0 & u & -v \\
-y & -u & 0 & w \\
-z & v & -w & 0
\end{array}
\right).
\end{equation}
In terms of these variables, the two lowest-order, Lorentz invariant forms
can be written as:
\begin{eqnarray}
\frac{1}{2} \theta_{\mu\nu}\theta^{\mu\nu} &=& - x^2 - y^2 - z^2 + u^2 + v^2 + w^2 \, ,
\nonumber \\
\frac{1}{8}\epsilon_{\mu\nu\alpha\beta}\theta^{\mu\nu}\theta^{\alpha\beta} &=& x\,w
 + y\,v + z\,u \,\,\, .
\label{eq:uvwxyz}
\end{eqnarray}
By inspection, the right-hand sides of Eq.~(\ref{eq:uvwxyz}) take a simple form if we
redefine our variables again, in terms of two sets of spherical polar coordinates:
\begin{eqnarray}
(x,y,z) \to (r_1, \theta_1, \phi_1), \nonumber \\
(w,v,u) \to (r_2, \theta_2, \phi_2).
\end{eqnarray}
The function $W(\theta)_{bc}$ can now be written
\begin{equation}
W(\theta)_{bc} = {\rm exp}(-c|r_1^2-r_2^2|) \, {\rm exp}(-b|\vec{r_1} \cdot \vec{r_2}|).
\label{eq:wbcr1r2}
\end{equation}
The $d^6\theta \equiv d^3 r_1 d^3 r_2$ integration is now easy to evaluate since Eq.~(\ref{eq:wbcr1r2})
depends only on $r_1 \equiv |\vec{r_1}|$, $r_2 \equiv |\vec{r_2}|$ and the angle between $\vec{r_1}$
and $\vec{r_2}$.  Furthermore, one may do the $r_2$ integration first, letting the $\vec{r_1}$
direction define the $z$-axis, so that the angular integration simplifies to $\int d\Omega_1 d\Omega_2
= \int 8\pi^2 \cdot d\cos\theta_2$.  One then finds that
\begin{equation}
\int d^6\theta \, W(\theta)_{bc} = \frac{16\pi^2}{b} \int dr_1\, r_1 \int
dr_2\,r_2 \, \exp(-c|r_1^2-r_2^2|)\, [1 - \exp(-b r_1 r_2)] \,\,\,.
\label{eq:divform}
\end{equation}
The $\int dr_1 \, r_1 \int dr_2 \cdot r_2 \,\, \exp (-c|r_1^2-r_2^2|)$ integration is divergent,
while the rest is finite.  However, this contribution also scales as $1/b$, while the finite part
does not, suggesting an alternative starting point in which this divergence is cancelled
\begin{equation}
W(\theta) = N \left[W(\theta)_{ac} - \frac{b}{a} W(\theta)_{bc}\right] \,\,\,,
\end{equation}
or explicitly,
\begin{equation}
{\rm W}(\theta) = N \exp(-\frac{c}{2}|\theta_{\mu\nu}\theta^{\mu\nu}|) \left[\exp(-
\frac{a}{8}|\epsilon_{\mu\nu\alpha\beta}\theta^{\mu\nu}\theta^{\alpha\beta}|)
- \frac{b}{a}\, \exp(-\frac{b}{8}|\epsilon_{\mu\nu\alpha\beta}\theta^{\mu\nu}\theta^{\alpha
\beta}|)\right] \,\,\, ,
\label{eq:ourw}
\end{equation}
where $N$ is a normalization constant and $a \neq b$.  The integral of this weighting function over all
$\theta$ is finite and can be performed analytically, yielding
\begin{eqnarray}
&& \int d^6\theta \,W(\theta) =  \nonumber \\
&& \frac{16\pi^2}{a} N \left[\frac{(2 c - b)}
{b (4 c^2 + b^2)}
+ \frac{2 b\,{\rm tanh}^{-1} \frac{(2 c - b)}{\sqrt{4 c^2 + b^2}}}
{(4 c^2+b^2)^{3/2}}
+\frac{2 b\,{\rm tanh}^{-1}\sqrt{\frac{b^2}{4 c^2+b^2}}}{(4 c^2+b^2)^{3/2}} \,\, - \,\,
(b \leftrightarrow a) \right].
\end{eqnarray}
Henceforth, we will use this result to choose a value for $N$ such that
\begin{equation}
\int d^6\theta \,W(\theta) = 1 \,\,\,.
\end{equation}
Although the form for $W(\theta)$ in Eq.~(\ref{eq:ourw}) is one of many possibilities, it is by far the
simplest, smooth function that meets our requirements.  Other choices that we have considered do not change
our field theoretic results qualitatively.  Therefore, in the Lorentz-invariant noncommutative theories
that follow, we adopt Eq.~(\ref{eq:ourw}) together with the simple parameter choice $a/2=b=c=\Lambda^4$, where
$\Lambda$ sets the scale of the new physics.
\section{Toy Models}\label{sec:three}
With a suitable weighting function defined, we now consider a few simple applications in toy
models.  These examples illustrate some of the qualitative features of the noncommutative vertex
modification.  We consider first noncommutative Yukawa theory, defined by the Lagrangian
\begin{equation}
{\cal L} = \int d^6 \theta \, W(\theta) \left[ \overline{\psi}\, (i \not\!\partial-m) \,\psi +
-\frac{1}{2} \phi\, (\partial_\mu \partial^\mu +m_\phi^2) \,\phi + \frac{\lambda}{2}(
\overline{\psi}\, \phi \star \psi + \mbox{h.c.})\right] \,\,\,,
\end{equation}
where $\phi$ is a real scalar field.  One star product has been removed from each term via integration
by parts and the discarding of surface terms.  The momentum-space Feynman rule for the fermion-fermion-scalar
vertex is
\begin{equation}
i \lambda \, \int d^6 \theta \, W(\theta) \exp[\frac{i}{2} p \cdot \theta \cdot q]
\label{eq:initv}
\end{equation}
where $p$ and $q$ are the four-momenta of the incoming and outgoing fermion.  Using the same notation
introduced in Section~\ref{sec:two}, we may write the vertex as
\begin{equation}
\int d^3 r_1 d^3 r_2 \, W(\vec{r}_1,\vec{r}_2) \exp\left[ i \vec{A} \cdot \vec{r_1} + i \vec{B} \cdot \vec{r_2}
\right] \,\,\, ,
\label{eq:vertex1}
\end{equation}
where $\vec{A}$ and $\vec{B}$ are functions of the four-momenta given by
\begin{equation}
\vec{A}= (p_0 \, \vec{q} - q_0 \, \vec{p})/2 \,\,\,\,\,\mbox{ and }
\,\,\,\,\, \vec{B}=(\vec{p}\times \vec{q})/2  \,\,\,.
\label{eq:aandb}
\end{equation}
Equation~(\ref{eq:vertex1}) is symmetric under the interchange of $\vec{A}$ and $\vec{B}$, and reduces
to the commutative result, $i\lambda$, in the limit where both are vanishing.  We will see, in every
application to be discussed in this paper, that either $\vec{A}$ or $\vec{B}$ may be set to zero, by working in
an appropriately chosen Lorentz frame. For concreteness, we will set $\vec{B}=0$ and let $\xi \equiv |\vec{A}|$.
Then, using the form for $W(\theta)$ in Eq.~(\ref{eq:ourw}), one may express the fermion-fermion-scalar vertex as
$i \lambda\, I_v(\xi)$, where
\begin{equation}
I_v(\xi)= -\frac{16 \pi^2 N}{a \xi} \int_0^\infty dr_1 \int_0^{r_1} dr_2\, [r_2 \sin(\xi r_1)+r_1 \sin(\xi r_2)]
\,(e^{-a r_1 r_2}-e^{-b r_1 r_2})\, e^{-c(r_1^2-r_2^2)} \,\,\,\,.
\label{eq:ivxi}
\end{equation}
This integral is finite.  One can see that $I_v(0)=1$ and $I_v(\infty)=0$ by using the fact that
\begin{equation}
\lim_{\xi\rightarrow 0} \frac{\sin(\xi r_i)}{\xi} = r_i \,\,\,\,\, \mbox{ and }\,\,\,\,\,
\lim_{\xi\rightarrow \infty} \frac{\sin(\xi r_i)}{\pi r_i}=\delta(r_i) \,\,\,.
\end{equation}
The integral may be done analytically, though the precise form of the result (given in the appendix) is not
particularly enlightening.  Qualitatively, $I_v(\xi)$ is a function with finite area that drops off quickly as
$\sqrt{\xi}$ exceeds the typical scale of noncommutativity, which is set by the choice of the parameters $a$,
$b$, and $c$.  In fact, this result could be anticipated from Eq.~(\ref{eq:initv}), which has the form of a
six-dimensional Fourier transform of the weighting function, {\em i.e.} $\widetilde{W}(\omega_{\mu\nu})$, where
the variable conjugate to $\theta$ takes a particular value, $\omega^{\mu\nu}=(p^\mu q^\nu-q^\mu p^\nu)/4$.  It
is not hard to see that the integral of $\widetilde{W}(\omega_{\mu\nu})$ over all $\omega$ is simply $W(0)$,
which is finite. Thus, $\widetilde{W}$ is a function that we would expect to drop off at large values of the
argument.

One might wonder at this point whether it would have been easier to specify the unknown function $\widetilde{W}$
in the conjugate space from the start.  From a model building perspective, such an approach would be contrived.
We have made the reasonable aesthetic choice that the function appearing directly in the Lagrangian
Eq.~(\ref{eq:effL}) be taken as simple as possible.  One might hope that this will make an eventual physical
interpretation of $W(\theta)$ more transparent.  In addition, our weighting function is more easily applied to
noncommutative theories, such as those requiring an expansion in $\theta$, that have a more complicated vertex
structure than in Eq.~(\ref{eq:initv}). In any case, the qualitative features of our results depend only on the
fact that the weighting function falls off at a well-defined scale.

\begin{figure}[t]
\epsfxsize 3.3 in \epsfbox{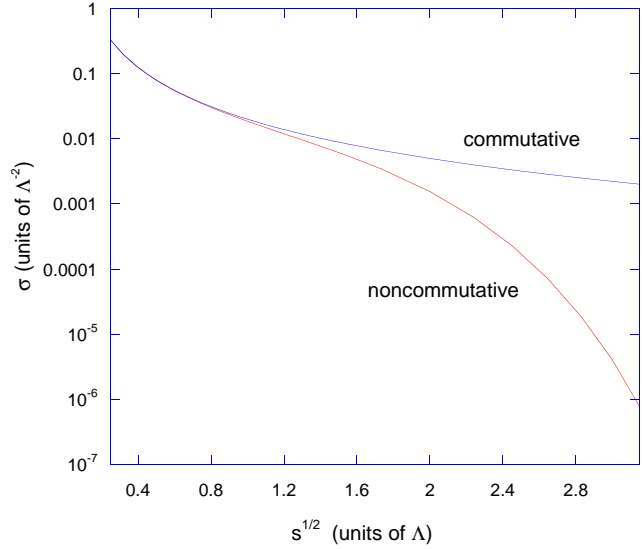} \caption{Difermion production in the
Yukawa theory, for the canonical parameter choice $a/2=b=c=\Lambda^4$, and negligible
fermion masses.}
\label{fig:ysct}
\end{figure}

The simplest process we can study in Yukawa theory is the scattering
$f_1 \bar{f}_1 \rightarrow f_2 \bar{f}_2$, for two distinct types of fermions.
This is purely an $s$-channel process; the two vertices will have noncommutative
exponential factors that depend on the factors $p_1 \cdot \theta \cdot \bar{p}_1$ and
$p_2 \cdot \theta \cdot \bar{p}_2$.  Working in the center-of-mass frame, the spatial
vectors $\vec{p}_i=-\vec{\bar{p}}_i$ and the quantity that we called $\vec{B}$ in Eq.~(\ref{eq:aandb})
will vanish.  The spin-averaged differential cross section is isotropic, and the full scattering cross
section is given by
\begin{equation}
\sigma = \frac{\lambda^4}{16 \pi} \, I^4_v(\xi_s)
\frac{(s-4 m^2)^2}{s [(s-m_\phi^2)^2 + m_\phi^2 \Gamma_\phi^2]} \,\,\,,
\end{equation}
where
\begin{equation}
\xi_s = \frac{1}{4} \sqrt{s} (s-4 m^2)^{1/2} \,\,\,\mbox{ and } \,\,\,
\Gamma_\phi = \frac{8 \lambda^2}{\pi m_\phi^5} \xi^3_{m_\phi^2} \, I_v(\xi_{m_\phi^2}) \,\,\,.
\end{equation}
For $\sqrt{s}$ larger than the typical energy scale set by the parameters $a$, $b$ and $c$ in
$I_v$, the cross section is suppressed relative to the commutative result.  This behavior is
illustrated in Fig.~\ref{fig:ysct}.

The modification to the differential cross section is less trivial if one considers processes
involving fermions of the same type in the initial and final state.  In this case, there are
both $s$- and $t$-channel contributions to the amplitude that receive {\em different}
noncommutative corrections.  Rather than examining $f \bar{f} \rightarrow f \bar{f}$
scattering in the present example, we defer this discussion to the next section where we
study Bhabha scattering in QED, which presents a nearly identical calculation.

\begin{figure}[t]
\epsfxsize 3.3 in \epsfbox{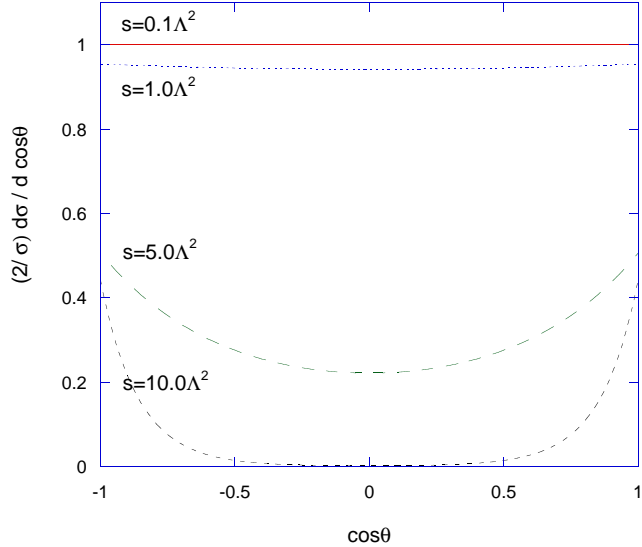} \caption{Differential cross section for two-into-two scattering
in $\phi^4$ theory, for the canonical parameter choice $a/2=b=c=\Lambda^4$.}
\label{fig:p4dcs}
\end{figure}

Instead, to illustrate the dramatic effects that the noncommutative vertex modification
can have on angular distributions, we consider tree-level scattering in $\phi^4$ theory,
defined by the Lagrangian
\begin{equation}
{\cal L} = \int d^6 \theta \, W(\theta) \, \left[
-\frac{1}{2} \phi\, (\partial_\mu \partial^\mu +m_\phi^2) \,\phi - \frac{\lambda}{4!}\,
(\phi \star \phi)\,(\phi \star \phi) \right] \,\,\,.
\end{equation}
One star product has been removed from the interaction term via integration by parts.
Consider tree-level, $\phi\phi\rightarrow\phi\phi$ scattering in the center-of-mass frame:
no example could be simpler in the commutative limit.  In the noncommutative case, the
momentum-space Feynman rule for the $\phi^4$ vertex has the form
\begin{equation}
i {\cal M} = 2 \cdot \frac{\lambda}{4!} \int d^6 \theta \, W(\theta) \, \left[
\exp[-\frac{i}{2}(p_1 \cdot \theta \cdot p_2 + p_3 \cdot \theta \cdot p_4)]
+ \mbox{ 11 other permutations}) \right]\,\,\,,
\label{eq:p4v1}
\end{equation}
where the momenta are outwardly directed.  Now assume that $p_1$ and $p_2$ correspond to
incident particles with beam energy $E$ and three-momentum magnitude $p$.  One can
show that each of the twelve terms in Eq.~(\ref{eq:p4v1}) can be written in the same
form as Eq.~(\ref{eq:vertex1}), with either $\vec{A}$ or $\vec{B}$ identically zero.
Thus, using the symmetry of the vertex function under $\vec{A} \leftrightarrow \vec{B}$,
one can express each in terms of the function $I_v(\xi)$, defined in
Eq.~(\ref{eq:ivxi}), for some choice of argument $\xi$.  We find
\begin{equation}
i {\cal M} = i \frac{\lambda}{3} \left[ I_v(\xi_1)+I_v(\xi_2) + I_v (\xi_3)\right]\,\,,
\end{equation}
where
\begin{eqnarray}
\xi_1 &=& \sqrt{2} E p (1+\cos\theta_0)^{1/2}\,, \nonumber \\
\xi_2&=&\sqrt{2} E p (1-\cos\theta_0)^{1/2}\,, \nonumber \\
\xi_3&=&p^2 \sin\theta_0\,,
\end{eqnarray}
and where $\theta_0$ is the center-of-mass scattering angle.  This leads immediately
to the differential cross section
\begin{eqnarray}
\frac{d\sigma}{d c_{\theta_0}}&=&\frac{\lambda^2}{288 \pi s} \left[
I_v \left( \frac{\sqrt{s}}{2\sqrt{2}} (s-4 m_\phi^2)^{1/2} (1+ c_{\theta_0})^{1/2}\right)
+
I_v \left( \frac{\sqrt{s}}{2\sqrt{2}} (s-4 m_\phi^2)^{1/2} (1- c_{\theta_0})^{1/2}\right)
\nonumber \right.\\
&& \left. + I_v \left( \frac{1}{4} (s-4 m_\phi^2) (1- c^2_{\theta_0})^{1/2}\right) \right]^2 ,
\label{eq:dcres}
\end{eqnarray}
with $c_{\theta_0} \equiv \cos\theta_0$. This result reduces to the commutative one at threshold, where the
argument of $I_v$ vanishes in each of the three terms.  The behavior of this differential cross section
for a variety of $\sqrt{s}$ is shown in Fig.~\ref{fig:p4dcs}.
The differential cross section is normalized to half the total cross section, which is the
value of $d\sigma/d \cos\theta$ at any angle in the limit where the amplitude becomes isotropic.
The form of the results in Fig.~\ref{fig:p4dcs} can be understood by noting that only
when $\cos\theta=\pm 1$ are there contributions to the differential cross section
that are unsuppressed in the large $\sqrt{s}$ limit.  The effect of the tree-level
noncommutative  vertex modification is therefore to suppress dramatically $\phi\phi$ production
transverse to the beam direction.  We will see similar modifications to differential cross
sections in the more realistic examples to be considered in the next section.

\section{NCQED}\label{sec:four}

We now consider the application of our approach to noncommutative QED.  As we mentioned earlier, we
focus on the original proposal of Hayakawa~\cite{haya}, which has been studied as a realistic
phenomenological theory by a number of authors~\cite{collider}.   The Lagrangian is given by
\begin{equation}
\mathcal{L}=\int d^{6} \theta \, W(\theta) \, \left[\bar{\psi}
(i\not\!{\partial}-m) \psi - \frac{1}{4}F_{\mu\nu} F^{\mu\nu} +
\frac{e}{2} (\bar{\psi} \not\!\!{A}\star \psi + \mbox{h.c.})\right],
\label{eq:QED1}
\end{equation}
where, as in Sec.~\ref{sec:three}, we have removed one star product via integration by parts and the
discarding of surface terms.   In this theory, the noncommutative field strength is given by
\begin{equation}
F_{\mu\nu}=\partial_{\mu}A_{\nu}-\partial_{\nu}A_{\mu}-ie
[A_{\mu}\stackrel{\star}{,}A_{\nu}].
\label{eq:QED3}
\end{equation}
Aside from the modification to the ordinary two-fermion-one-photon vertex, this theory has photon
self interactions.  Unlike the Hayakawa formulation, however, the three-photon vertex is absent.
This interaction is proportional to $\sin(p_i\cdot\theta\cdot p_j/2)$, where the $p_i$ are external momenta,
but vanishes upon integration against the weighting function $W(\theta)$, which is an even function of
$\theta^{\mu\nu}$.  Here we focus on dilepton production, Bhabha scattering and diphoton production,
processes which are unaffected by the new four-photon vertex at lowest order.

We may write the Feynman rule for the two-fermion-one-photon vertex as
\begin{eqnarray}
i e \gamma^\mu \int d^{6} \, \theta W(\theta) \, \exp(\frac{i}{2} \, p_1
\cdot \theta \cdot p_2) \equiv i e \gamma^\mu I(p_1,p_2),
\label{eq:feynrule1}
\end{eqnarray}
where $p_1$ and $p_2$ are the momenta of the incoming and outgoing fermions, respectively.
As we will see below, the function $I(p_1,p_2)$ may be re-expressed in terms of the vertex integral $I_v(\xi)$
given in Eq.~(\ref{eq:ivxi}), for an argument $\xi$ that depends on the process in question.

\subsection{\label{sec:dilepton} Dilepton Production, $e^+e^-\to l^+l^-$}

The matrix element for Bhabha scattering that follows from Eq.~(\ref{eq:feynrule1}) is:
\begin{eqnarray}
i \mathcal{M} = e^2 \bar u(p_3) \, i\gamma^\nu { I}(p_3,p_4) \, v(p_4)
\frac{-ig_{\mu\nu}}{q^2+i\epsilon} \bar v(p_2) \, i\gamma^\mu {I}(p_1,p_2) \, u(p_1)
\nonumber \\
- e^2 \bar v(p_2) \, i\gamma^\nu { I}(p_2,p_4) \, v(p_4)
\frac{-ig_{\mu\nu}}{q'^2+i\epsilon} \bar u(p_3) \, i\gamma^\mu {I}(p_1,p_3) \, u(p_1).
\label{eq:mat_el_dilepton}
\end{eqnarray}
Here, $u$ and $v$ represent the momentum-space spinor wave functions for the
fermions and antifermions, respectively, and the propagator momenta are given by
$q=p_1+p_2$ and $q'=p_1-p_3$. The four vertex functions $I(p_i,p_j)$ can be evaluated using the
same approach presented in Sec.~\ref{sec:three}.  The functions $I(p_1,p_2)$ and $I(p_3,p_4)$ are
straightforward to simplify since the incoming and outgoing particles are travelling in opposite
directions in the center-of-mass frame, so that
$\vec{B}=0$ and $\vec{A}=s/4$.  Hence,
\begin{equation}
I(p_1,p_2) = I(p_3,p_4) = I_v(s/4) \,\,\,,
\end{equation}
where $s$ is the usual Mandelstam variable.  Here, and throughout this section, we ignore the fermion masses,
which are entirely irrelevant given the collider energies of interest (for example, those of a TeV-scale linear
collider).  Evaluating the remaining functions, $I(p_1,p_3)$ and $I(p_2,p_4)$, requires a little more effort.
Both are Lorentz-invariant functions of the given momenta, so that the result of the integration does not
depend on the frame in which it is performed.  For example, we can equate $I(p_1,p_3)$ to an integral
evaluated in a frame where $\vec{B}=0$, simplifying the calculation.  In this case, we can work in the $p_1$
rest frame, which is related to the center-of-mass frame by a boost
\begin{equation}
p_1 = \left(
\begin{array}{c}
E \\ 0 \\ 0 \\ p
\end{array}
\right) \longrightarrow
\Lambda \, p_1 = \left(
\begin{array}{c}
m \\ 0 \\ 0 \\0
\end{array}
\right) \,\,\,,
\end{equation}
where
\begin{equation}
\Lambda = \left(
\begin{array}{cccc}
\gamma & 0 & 0 & -\beta \gamma \\
0 & 1 & 0 & 0 \\
0 & 0 & 1 & 0 \\
-\beta\gamma & 0 & 0 & \gamma
\end{array}
\right),
\end{equation}
where $\gamma= (1-\beta^2)^{-1/2}$ with $\beta=p/E$.  On the other hand,
$\Lambda p_3$ has the spatial components $[0, p\,\sin\theta_0, - \gamma p\,(1-\cos\theta_0)]$, where
$\theta_0$ is the center-of-mass scattering angle.  After taking the limit $m \to 0$, the two vectors
$\vec{A}$ and $\vec{B}$ become $[0, 0, -E^2(1-\cos\theta_0)/2]=[0, 0, t/4]$ and $[0,0,0]$, respectively.  Hence,
\begin{equation}
I(p_1,p_3) = I(p_2,p_4) = I_v(t/4) \,\,\,.
\label{eq:ivt}
\end{equation}
Note that the simplification of $I(p_2,p_4)$ is accomplished using the approach just described, but
by evaluating the integral instead in the $p_2$ rest frame.

Squaring the matrix element and summing (averaging) over the final (initial)
fermion spin states gives
\begin{equation}
\overline{|\mathcal{M}|^{2}} =
2 e^4 \left[ I_v^4(s/4) (\frac{t^2 + u^2}{s^2}) +
2 \, I_v^2(s/4) I_v^2(t/4) \frac{u^2}{st} +
I_v^4(t/4) (\frac{u^2 + s^2}{t^2})\right], \label{eq:msqr1}
\label{eq:a2bhabha}
\end{equation}
where the Mandelstam variables are defined by
\begin{equation}
s = (p_1 + p_2)^2,  \,\,\,\,\,\,\,  t = (p_1 - p_3)^2,  \,\,\,\,\,\,\, u = (p_1 - p_4)^2 \,\,\, .
\end{equation}
The same results for $e^+e^-\to\mu^+\mu^-$ can be obtained easily by discarding
the $t$-channel contribution in Eq.~(\ref{eq:a2bhabha}), again ignoring all fermion
masses.  The spin-averaged, squared matrix element is:
\begin{equation}
\overline{|\mathcal{M}|^{2}} = 2 e^4 I_v^4(s/4) (\frac{t^2 +
u^2}{s^2}). \label{eq:msqr3}
\label{eq:mpmm}
\end{equation}
The differential cross section for both processes can be obtained by
multiplying the results above by the proper phase space factor:
\begin{equation}
\frac{d\sigma}{d \cos\theta_0} =
\frac{\overline{|\mathcal{M}|^{2}}}{32\pi s}  \,\,\, .
\label{eq:genform}
\end{equation}
Eqs.~(\ref{eq:mpmm}) and (\ref{eq:genform}) allow us quickly to place a bound on the noncommutativity scale.
Noting that at the highest LEP II energies ($\sim 189$~GeV), the cross section for dimuon production agrees
with the stardard model prediction up to $5$\% experimental error bars~\cite{CKN}, it is reasonable to require
that $I_v^4(s/4) \gtrsim 0.9$ which leads to the bound
\begin{equation}
\Lambda > 172\mbox{ GeV}\,\,\,\,\,\,\mbox{95\% C.L.}
\end{equation}
This is consistent with the bound on $\Lambda_{NC}$ found in Ref.~\cite{CKN}, and quoted in
Eq.~(\ref{eq:cknbound}).  Note that we cannot adopt the same definition of the noncommutative scale as
Ref.~\cite{CKN} because Eq.~(\ref{eq:lncdef}) is vanishing for the present choice of weighting function.

\subsection{\label{sec:diphoton}Diphoton Production, $e^+e^-\to\gamma\gamma$}

\begin{figure}[t]
\epsfxsize 3.3 in \epsfbox{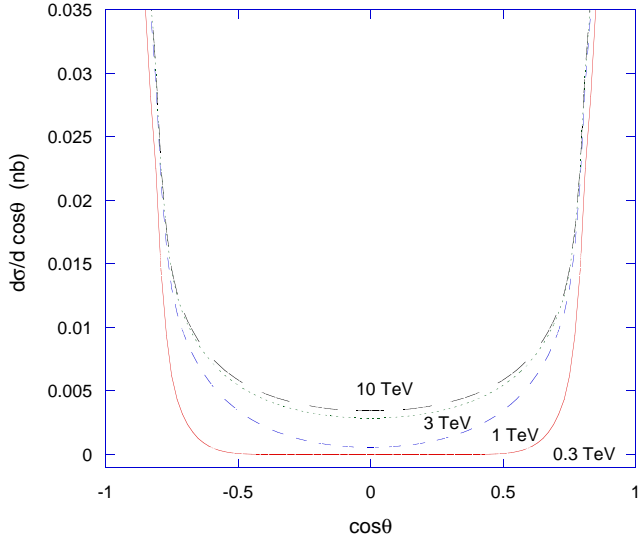} \caption{Differential cross section for diphoton production at
a linear collider with $\sqrt{s}=1.5$~TeV.  The curves shown correspond to the canonical parameter choice
$a/2=b=c=\Lambda^4$, with $\Lambda=0.1$, $1$, $3$ and $10$~TeV.}
\label{fig:dphot}
\end{figure}

Diphoton production is particularly interesting in this model since it is different from
canonical noncommutative theories (there is no three-photon vertex) and from the theory described in
Ref.~\cite{CKN} (there is no two-photon-two-fermion vertex).  The matrix element for diphoton production
can be written as:
\begin{eqnarray}
i \mathcal{M} = e^2 \bar v(p_2) \, i\gamma^\mu I(p_2,p_4) \,
\frac{i(\not\!q+m)}{q^2-m^2+i\epsilon} \, i\gamma^\nu I(p_1,p_3) \,
u(p_1) \, \epsilon_\mu^*(p_4) \, \epsilon_\nu^*(p_3) \nonumber \\
+ e^2 \bar v(p_2) \, i\gamma^\mu I(p_2,p_3) \,
\frac{i(\not\!q'+m)}{q'^2-m^2+i\epsilon} \, i\gamma^\nu I(p_1,p_4) \,
u(p_1) \, \epsilon_\mu^*(p_3) \, \epsilon_\nu^*(p_4).
\label{eq:mat_el_diphoton}
\end{eqnarray}
Here, $p_1$ ($p_2$) is the momentum of the incoming fermion (anti-fermion), while $p_3$ and $p_4$ are
the photon momenta; $u$ and $v$ are the momentum-space spinor wave-functions for the fermion
and antifermion, $\epsilon_\mu$ represents a photon polarization vector and the propagator
momenta are given by $q=p_1-p_3$ and $q'=p_1-p_4$.  Again we may simplify the factors of $I(p_i,p_j)$
by boosting to appropriate Lorentz frames.  One finds, in addition to Eq.~(\ref{eq:ivt}), that
\begin{equation}
I(p_1,p_4)=I(p_2,p_3)=I_v(u/4) \,\,\,,
\end{equation}
from which it follows that
\begin{equation}
\frac{d\sigma}{d\cos\theta_0} =
\frac{\pi \alpha}{s} \left[I_v^4(t/4) (\frac{u}{t}) +
I_v^4(u/4) (\frac{t}{u})\right].
\end{equation}
This cross section shows the same suppression in directions transverse to the beam that we encountered
in the case of $\phi^4$ theory.   This is displayed in Fig.~\ref{fig:dphot}, for a linear $e^+ e^-$
collider with $\sqrt{s}=1.5$~TeV, for a range of noncommutative scales $\Lambda$.   The standard model
result is recovered in the large $\Lambda$ limit, and the figure shows discernable deviations that
could, in principle, be extracted via a fit to the eventual data.

\subsection{Coulomb Potential}

It is interesting to note that the high-momentum modification of the photon-fermion-fermion
vertex alters the Coulomb potential at short distances.  We evaluate the potential by studying the
non-relativistic reduction of the $t$-channel scattering amplitude of two distinguishable fermions.
For each vertex, $p_i \cdot \theta \cdot p_f$ is such that $\xi =|\vec{A}|= m |\vec{q}| /2$ and
$|\vec{B}|=|\vec{p}_i \times \vec{q}|$, where $\vec{q}$ is the spatial momentum transfer.  We therefore
may discard $\vec{B}$ in the nonrelativistic reduction, since it is higher order in $p_i/m \ll 1$. In momentum
space, the usual form for the nonrelativistic potential is modified by an additional factor of $I_v^2(m\, q/2)$:
\begin{equation}
V(r) =  \frac{e^2}{4 \pi^2 \, i\,  r} \int_{-\infty}^{\infty} dq \,
\frac{1}{q}\exp(i\,q\,r) \, I_v^2(m\, q/2) \,\,\,.
\label{eq:modpot}
\end{equation}
Normally, one completes this calculation by performing a complex contour integration, with the contour closed
in the upper half plane.  In this case, one cannot argue that the contour at infinity has a vanishing
contribution to the result, due to the additional function in the integrand.  It is straightforward to
evaluate this integral numerically.  Here we will simply state the important qualitative features.  At large
distances, the Coulomb potential is not altered; this region corresponds to small $q$ for which
$I_v^2(m\, q/2)\approx 1$.  On the other hand, as one approaches the origin, one finds that $V(r)$ remains
nonsingular.  Writing Eq.~(\ref{eq:modpot}) as
\begin{equation}
V(r) =  \frac{e^2}{4 \pi^2 \,  r} \int_{-\infty}^{\infty} dq \, \frac{\sin q\,r}{q} \, I_v^2(m\,q/2) \,\,\,,
\label{eq:modpot2}
\end{equation}
one may use the fact that $\lim_{r\rightarrow 0} \sin q\,r /r = q$ to see that
\begin{equation}
V(0) = \frac{e^2}{4 \pi^2} \int_{-\infty}^{\infty} dq\, I_v^2(m\, q/2) = \frac{2 \alpha}{\pi m}
\int_{-\infty}^{\infty} d\xi\,I_v^2(\xi)  \,\,\,.
\end{equation}
The last integral is finite and is ${\cal O}(\Lambda^2 \alpha/(\pi m))$, for our parameter choice
$a/2=b=c=\Lambda^4$.  Hence, the singularity at the origin is not present, a result that one might attribute
to the fuzziness of the noncommutative space. It is important to note that this result is independent of the
detailed form of $W(\theta)$ and would be obtained for any weighting function with finite volume in momentum
space, since such a function provides an ${\cal O}(\Lambda)$ ultraviolet cut-off in Eq.~(\ref{eq:modpot}).

\section{Four-fermion Interactions}\label{sec:five}

In the theories that we have considered, the Lorentz-invariant, noncommutative modification to the
tree-level vertices leads to a suppression in scattering cross sections at energies above the typical
scale of the new physics.  It is not unreasonable to guess that, at least in some theories, a violation
of unitarity in the commutative limit may be cured by the additional high-momentum suppression of the
noncommutative vertex.  In this subsection, we will give one example where this is the case, working
at tree-level.

Consider a theory of two distinguishable fermions, $\psi_1$ and $\psi_2$, with the nonrenormalizable
interaction Lagrangian
\begin{equation}
{\cal L}_{int} = \frac{1}{M^2} (\bar{\psi_1}\star \psi_1)(\bar{\psi_2}\star \psi_2) \,\,\,.
\label{eq:dimsix}
\end{equation}
We may assume that the fermions have masses $m_1$ and $m_2$, though we will only be interested in the
behavior of the scattering cross sections at high energies, where these masses are irrelevant.

The possible tree-level, two-into-two scattering processes are fermion-anti-fermion annihilation,
and the scattering of fermions and/or antifermions of a different type.  With the interaction
written as in Eq.~(\ref{eq:dimsix}) (again with one star product removed), the relevant differential
cross sections in the center-of-mass frame and in the high-energy limit are given by
\begin{equation}
\frac{d\sigma}{dc_{\theta_0}}( \psi_i \bar\psi_i \rightarrow \psi_j \bar\psi_j) = \frac{s}{32 \pi \, M^4} I_v^2
\left[s(1-c_{\theta_0})^{1/2}/(2\sqrt{2})\right]  \,\,\,,
\end{equation}
\begin{equation}
\frac{d\sigma}{dc_{\theta_0}}( \psi_i \psi_j \rightarrow \psi_i \psi_j) = \frac{s\, (1-c_{\theta_0})^2}{128 \pi M^4}
I_v^2\left[ s (1-c^2_{\theta_0})^{1/2}/4\right] \,\,\,,
\end{equation}
and
\begin{equation}
\frac{d\sigma}{dc_{\theta_0}}( \psi_i \bar{\psi}_j \rightarrow \psi_i \bar{\psi}_j) =
\frac{s\, (1-c_{\theta_0})^2}{128 \pi M^4}
I_v^2\left[ s (1-c_{\theta_0})^{1/2}/(2\sqrt{2})\right] \,\,\,,
\end{equation}
for $i\neq j$.  Clearly, for any $\theta_0 \neq 0 \mbox{ or }\pi$, these differential cross sections
are suppressed in the large $\sqrt{s}$ limit:  the exponential damping of $I_v$ above the noncommutative scale
wins over the $s/M^4$ growth of the commutative result.  This conclusion remains unchanged when one integrates
over the center-of-mass scattering angle $\theta_0$.  Provided that the scale of noncommutativity is not far in
the ultraviolet, one can prevent an unlimited growth in the total cross section that would exceed the
unitarity bounds.  To quantify this, one may consider the partial wave decomposition
\begin{equation}
\left(2 \frac{d\sigma}{d\Omega}\right)^{1/2} = \frac{1}{\sqrt{s}} \sum_{J=0}^\infty
(2 J +1) P_J(\cos\theta_0) M_J
\end{equation}
where the partial wave amplitudes must satisfy the unitarity constraint $|M_J| < 1$~\cite{quigg}.  In all of
the examples above, we find that $s$-wave unitarity provides the tightest bound, which numerically is consistent
with the inequality
\begin{equation}
M \gtrsim 0.4 \, \Lambda
\end{equation}
for our canonical parameter choice $a/2=b=c=\Lambda^4$.  Thus, if $M$ is not far below the scale of
noncommutative physics, one need not worry that this theory will violate unitarity at tree-level.

\section{Conclusions}\label{sec:six}
In this paper, we have revisited a proposal for constructing Lorentz-invariant, noncommutative
field theories.  Canonical noncommutative field theories involve a Lorentz-violating parameter,
while nature shows no indication that Lorentz-invariance is broken.  Experimental limits on the
amount of Lorentz violation that is tolerable in the low-energy effective theory force the typical
energy scale of the noncommutative interactions to be above the center-of-mass energies that can be
probed directly in planned collider experiments.  This motivated the proposal in Ref.~\cite{CCZ}
to formulate a Lorentz-conserving alternative to the canonical models.

As we have reviewed earlier, this proposal involves extending the coordinate algebra by promoting the
noncommutativity parameter of the canonical theories to a fictitious coordinate.  The mapping of this
algebra into a field theory is again accomplished through a star product, but the new coordinates
are integrated in the action with a weighting factor, which sets the scale of the noncommutative physics.
The resulting four-dimensional effective theory remains Lorentz invariant and a function of ordinary
coordinates only.  Unlike previous phenomenological work based on this approach, we have considered theories
in which the prescription described for constructing a Lorentz-invariant noncommutative theory could be
implemented and studied without recourse to a low-momentum expansion. In particular, this allows us to study
how cross sections behave as typical energies exceed the noncommutative scale, as we have done in both toy
models (Yukawa and $\phi^4$ theory) and the more realistic case of noncommutative QED for fields in the
lepton sector, where the matter have charges $\pm 1$ or $0$.

Interestingly, we find that cross sections drop quickly as the center-of-mass energy exceeds the
noncommutative scale.  Roughly speaking, the noncommutative vertex modification smears out the
point-like interaction so that at high center-of-mass energies the incoming scatterer sees less
and less of the target.  Equivalently, the prescription for integrating over the fictitious $\theta$
coordinates with the weighting function $W(\theta)$ introduces form-factors in the tree-level
vertices of the theory.  As we saw in a number of cases, this also has consequences for the angular
distributions of scattering processes, since the various contributions to a scattering amplitude with
differing angular dependence in the commutative limit may each receive different noncommutative
corrections.  The modifications to the QED processes of dilepton production, Bhabha scattering and
diphoton production were all presented.   While Ref.~\cite{CKN} evaluated the lowest-order
effects of noncommutativity on these processes in one version of noncommutative QED (which is
appropriate for determining the collider bounds on the model), the results presented here would be
useful for a qualitative comparison to the data if, for example, a $1.5$~TeV International Linear
Collider (ILC) were to discover and study noncommutative physics at a scale comparable to its
center-of-mass energy.

Finally, we noted that a modification to a tree-level interaction that suppresses a given
cross-section at high center-of-mass energies may alter one's conclusions as to whether the cross
section ever violates the constraints from unitarity of the $S$-matrix.  We demonstrated this in the
case of a simple four-fermion interaction.  Since our vertex modification introduces an exponential
suppression factor at high $\sqrt{s}$, then a theory whose cross section grows only as a power $\sqrt{s}$
in the commutative limit, will not grow indefinitely in the noncommutative case.  Using the constraint
from partial-wave unitarity, we obtained a bound on the coefficient of the four-fermion interaction
so that the theory would remain unitary for arbitrarily large $\sqrt{s}$.  Whether this could be an
avenue for constructing a new type of Higgsless theory is an interesting question, but one that first
requires a more tractable, all-orders formulation of noncommutative standard model.

\begin{acknowledgments}
We thank Josh Erlich for useful comments. We thank the NSF for support under Grant No.~PHY-0456525.
\end{acknowledgments}

\appendix
\section{}
\begin{figure}[t]
\epsfxsize 3.3 in \epsfbox{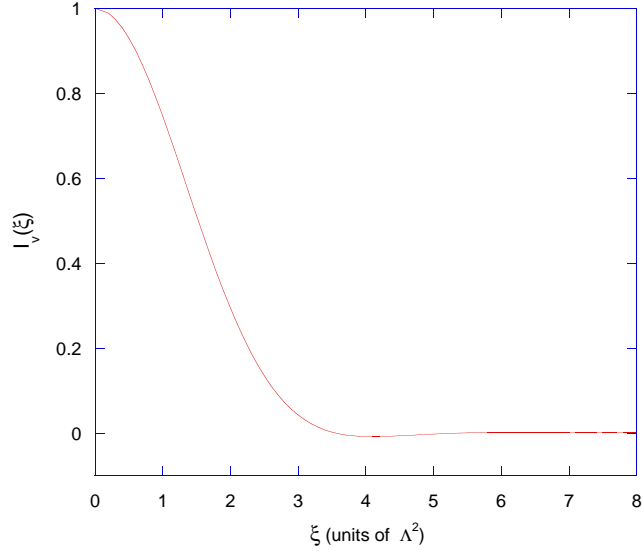} \caption{Plot of the vertex integral
function, for the canonical parameter choice $a/2=b=c=\Lambda^4$.}
\label{fig:thefnc}
\end{figure}
The integrals in Eq.~(\ref{eq:ivxi}) may be evaluated, yielding

\begin{eqnarray}
{\rm I}_v(\xi) &=& \frac{16\pi^2N}{a\xi} \Big[\frac{{\rm exp}
(\frac{-\xi^2}{4b}) \sqrt{\pi}
{\rm Erfi}(\frac{\xi}{2\sqrt{b}})} {2c\sqrt{b}} \nonumber  \\
&+& \frac{\xi}{c^2}
\sum\limits_0^\infty \frac{n+1}{2n+1}
\{\frac{(\frac{b}{2c})(1-\frac{b}{2c})^{2n+1}+(\frac{b}{2c})^{2n+2}}
{(1+\frac{b^2}{4c^2})^{2+n}}\} _1F_1[2+n, 3/2, -\xi^2/
(4c+b^2/c))] \nonumber \\
&&  - (b \leftrightarrow a)
\Big],
\end{eqnarray}
where Erfi is an imaginary error function, and $_1F_1$ is a confluent
Hypergeometric Function of the first kind.  This function is plotted in
Fig.~\ref{fig:thefnc}.


\end{document}